\documentclass[aps,prd,preprint,amsmath,amssymb,nofootinbib]{revtex4}

\usepackage{graphicx}

\begin{document}

\title{On the Factorisation of the Connected Prescription for Yang-Mills
Amplitudes}

\preprint{Saclay/SPhT-T06/188}

\author{C. Vergu}
\email{cvergu@cea.fr}
\affiliation{%
  Service de Physique Th\' eorique, CEA-Saclay\\
  F-91191 Gif-sur-Yvette CEDEX, France
}%

\begin{abstract}
  We examine factorisation in the connected prescription of Yang-Mills
  amplitudes.  The multi-particle pole is interpreted as coming from
  representing delta functions as meromorphic functions.  However, a
  naive evaluation does not give a correct result.  We give a simple
  prescription for the integration contour which does give the correct
  result.  We verify this prescription for a family of gauge-fixing
  conditions.
\end{abstract}

\maketitle

\section{Introduction}
\label{sec:intro}

It has been known for some time now that the gauge theory amplitudes
in perturbation theory are considerably simpler than the Feynman rules
would lead us to believe (see~\cite{Mangano:1990by} for a review).
This has been verified at tree~\cite{Parke:1986gb,Berends:1987me} and
loop~\cite{Bern:1996je} levels for ever-increasing numbers of external
lines and, in some particular cases such as the MHV (maximally
helicity violating) amplitudes, for arbitrary number of external
lines.  This simplicity is particularly striking for supersymmetric
theories but, in some cases, holds true for non-supersymmetric
theories.

There is as yet no satisfactory answer to why this is so.  After the early
work of Nair~\cite{Nair:1988bq}, Witten~\cite{Witten:2003nn} tried to
explain this simplicity by postulating a duality between the maximally
supersymmetric Yang-Mills theory and a string theory in (super-)twistor
space.  While the details of this duality are not yet fully understood, it
has already spurred some considerable
advances~\cite{Cachazo:2004kj,Britto:2005fq,Birthwright:2005ak,Risager:2005vk}
in computational techniques, especially at tree level.  For progress at the
loop level, see
refs.~\cite{Brandhuber:2005kd,Cachazo:2004zb,Bena:2004xu,Britto:2004nc,Britto:2005ha,Bern:2005cq,Berger:2006ci}.

Soon after Witten's breakthrough, it became
clear~\cite{Roiban:2004vt,Roiban:2004ka,Roiban:2004yf} that the amplitudes
can be computed from a subset of the contributions Witten proposed
initially (the connected instantons).  We will call this prescription for
computing Yang-Mills amplitudes `the connected prescription.'  The
connected prescription is very elegant, even if it is not the simplest
computationally.  However, it has the desirable feature of preserving (in a
manifest form) a large part of the (sometimes hidden) symmetries of the
scattering amplitudes.  A number of such arguments for the validity of this
prescription were presented in~\cite{Roiban:2004yf}, along with explicit
computations in some particular cases.  These non-trivial tests leave
little doubt that this prescription is correct.  For a review covering the
early work on the subject, see~\cite{Cachazo:2005ga}.

One major drawback of the connected prescription is that the
factorisation properties of the amplitudes are not obvious (in the
so-called completely disconnected prescription the factorisation
properties are obvious but the Lorentz invariance and parity are
obscured).

Some arguments that the connected and the completely disconnected
prescription are equivalent appeared in ref.~\cite{Gukov:2004ei}.  From
this point of view, the factorisation can be proven by proving the
equivalence with the completely disconnected prescription first.  Berkovits
and Motl also argued in~\cite{Berkovits:2004tx} that factorisation should
be a consequence of the possibility of formulating a string field theory
(SFT) which, so long as it is consistent, should have the right
factorisation properties.  This SFT is the off-shell extension of an
alternative string theory interpretation~\cite{Berkovits:2004hg} proposed
by Berkovits.

We believe that there should be a simple argument for the factorisation,
preferably one which will properly account for particles exchanged in
different channels.  This is difficult to achieve in this formalism because
the formalism is always on-shell, and an off-shell description is not
available.  Such understanding would be helpful in order to construct loop
amplitudes in this picture through use of unitarity.

The organisation of the paper is as follows.  First we briefly review
the connected prescription and factorisation.  We then give an
interpretation of the delta functions from the connected prescription
in terms of meromorphic functions.  Next, we discuss a scaling limit
inspired by Berkovits's string theory interpretation and its
consequences.  Finally, we argue that this implies factorisation.

\section{Review}
\label{sec:rev}

\subsection{Connected Prescription}
\label{sec:rev_conn}

In (super-)Yang-Mills theories, tree level $n$-point gluon scattering
amplitudes can be colour-decomposed as follows,
\begin{equation}
  \mathcal{A}_n(\{k_i, h_i, a_i\}) = g^{n-2} \sum_{\sigma \in
    S_n/\mathbb{Z}_n} Tr(T^{a_{\sigma(1)}} \cdots T^{a_{\sigma(1)}})
  A_n(\sigma(1), h_{\sigma(1)}; \ldots; \sigma(n), h_{\sigma(n)}),
\end{equation} where $k_i$ is the momentum of the $i$-th particle and $h_i$
is its helicity (we consider all particles to be out-going), $a_i$ label the
generators $T^{a_i}$ of the colour algebra, $g$ is the Yang-Mills coupling
constant, and $\sigma \in S_n/\mathbb{Z}_n$ instructs us to sum only over
cyclically non-equivalent permutations $\sigma$.  In the following we will
be interested only in the \emph{partial} amplitudes $A_n$.

In~\cite{Roiban:2004yf,Roiban:2004ka,Roiban:2004vt} a formula for
computing these partial amplitudes was presented. In a slightly
different notation, it reads
\begin{equation}
  \label{eq:connected}
  A_n = \int \frac{\prod_{k=0}^d d^{4|4} a_k^I \prod_{i=1}^n d
    \sigma_i}{\text{vol}(\text{Gl}(2))} \prod_{i=1}^n \frac
  {V_i(Z(\sigma_i))}{\sigma_i - \sigma_{i+1}},
\end{equation} where $I \in \{1,2,3,4|1',2',3',4'\}$, where the
external wavefunctions are,
\begin{equation}
  V_i(Z) = \int \frac {d \xi_i} {\xi_i} \delta^2(\pi_i - \xi_i
  \lambda) \exp (i \xi_i \left[\mu, \bar{\pi}_i\right])
  \exp (i \xi_i \psi^A \eta_{i A})
\end{equation} and where the twistor-space positions $Z=(\lambda, \mu |
\psi)$ are on curves parametrised via,
\begin{equation}
  Z^I(\sigma) = \sum_{k=0}^d a_k^I \sigma^k,
\end{equation} where $\sigma$ and $\xi$ are complex variables, and
where $\pi$ is a commuting spinor.

The $\text{vol}(\text{Gl}(2))$ arises from the $\text{Gl}(2)$ symmetry of
the integral, which renders it ill-defined.  In order to obtain a
meaningful result, we must pick a representative from each $\text{Gl}(2)$
orbit, that is fix a gauge.

Eq.~(\ref{eq:connected}) yields the scattering amplitude of on-shell
gluons with momenta $p_i^{\alpha \dot{\alpha}} = \pi_i^\alpha
\bar{\pi}_i^{\dot{\alpha}}$.  The helicities are obtained as the
coefficients of an expansion in the Grassmann variables $\eta_i$.  If
there's no factor of $\eta_i$ present, we take the helicity of $i$-th
particle to be $+$, and if there are four factors of $\eta_i$ present
($\eta_i^{1'} \eta_i^{2'} \eta_i^{3'} \eta_i^{4'}$) we take the
helicity of $i$-th particle to be $-$.  If $q$ is the number of
helicity minus particles, then $d = q - 1$.

One interesting point is that fact that, apart from the momentum conserving
delta functions and after gauge-fixing the $\text{Gl}(2)$ symmetry, the
number of integrals matches the number of unknowns.  Therefore, the
integrals above only receive contributions from isolated points.  A
puzzling fact, already recognised in ref.~\cite{Roiban:2004yf}, is that, in
order to obtain a correct result, complex solutions to the equations
imposed by the above delta functions above had to be included.

The inclusion of these complex solutions is quite unnatural from the
point of view of the delta functions and therefore the position in
ref.~\cite{Roiban:2004yf} was to not consider the above integrals as
`real' integrals, but just as a notation for the procedure of summing
a certain `Jacobian' over the roots of the equations imposed by the
delta functions.

However, it turns out that this interpretation of eq.~(\ref{eq:connected})
is not very useful for displaying the factorisation properties.  We will
give below an interpretation of these delta functions as meromorphic
functions which is suited for proving factorisation.

\subsection{Factorisation}
\label{sec:rev_fact}

The partial amplitudes defined above can only have poles when the sum of
more than two adjacent momenta goes on-shell.  More precisely, if $P
\equiv p_1 + \cdots p_m$ and $P^2 \rightarrow 0$, then we have
\begin{equation}
  \label{eq:fact}
  A_n(p_1, \ldots, p_n) \sim \sum_{h=\pm} A_{m+1}(p_1, \ldots, p_m, P^h)
  \frac i {P^2} A_{n-m+1}(P^{-h}, p_{m+1}, \ldots, p_n),
\end{equation} where $h = \pm$ represents the sum over the two
helicities of the internal factorized particle

\section{Delta Functions}
\label{sec:delta}

We now come the the question of how to interpret the delta functions
from the connected prescription.  The defining property of a delta
function is the following property
\begin{equation}
  f(a) = \int d x \delta(x - a) f(x),
\end{equation} for all functions $f$.

We propose to take $\delta(z - z_0) \equiv \frac 1 {2 \pi i} \frac 1
{z - z_0}$.  Also, the integral should be interpreted as a contour
integral along a contour which encircles the point $z_0$ in the
complex $z$ plane.  Then, the defining property of the delta function
results from Cauchy's theorem (assuming that $f$ has no poles inside
the integration contour).

All the delta functions functions which appear in the connected
prescription can be interpreted in this way.  This interpretation is
compatible with the usual properties of Fourier integrals if we define the
Fourier integral to be a complex integral along a contour from zero to
infinity, chosen in such a way to insure the convergence.  For example, in
the case of real $z$, the Fourier transform of the identity is defined as
follows\footnote{The minus sign might seem a bit strange, but it will be
  without consequence for our calculation since the phase of the amplitudes
  in the connected prescription is ambiguous anyway.}
\begin{equation}
  \int_0^{+i \infty} \frac {d k}{2 \pi} e^{i k z} = - \frac 1 {2 \pi
    i} \frac 1 z = - \delta(z).
\end{equation}  This kind of contour, from zero to infinity was
already used in~\cite{Cachazo:2004kj} in a heuristic discussion of
twistor-space propagator.

This interpretation is fully compatible with the delta functions
manipulations in ref.~\cite{Roiban:2004yf}.  For example, we have
\begin{equation}
  \int \delta(f(z)) = \frac 1 {2 \pi i} \oint \frac {d z}{f(z)} =
  \sum_{z_i \in \{z| f(z)=0\}} \frac 1 {f'(z_i)}.
\end{equation}

Note also that, multiple roots of $f(z) = 0$ do not contribute and the
result is obtained by using the Jacobian itself (and not the absolute
value of the Jacobian).  This is indeed what is required to obtain a
correct answer in the connected-prescription
computation~\cite{Roiban:2004yf}.

In what follows, we will leave implicit the replacement of delta
functions with their interpretation as meromorphic functions described
above.

\section{Factorisation}
\label{sec:fact}

A heuristic picture of when factorisation occurs is clearest in
Berkovits's string.  Our discussion below will not depend however on
the details of the Berkovits model.  An amplitude factorises when the
world-sheet (which is topologically a disk for tree amplitudes) can
be pinched, separating the vertex operators into two sets.  This can be
pictured as two disks joined by a very long, thin strip.

\begin{figure}
  \centering
  \includegraphics[width=.75\textwidth]{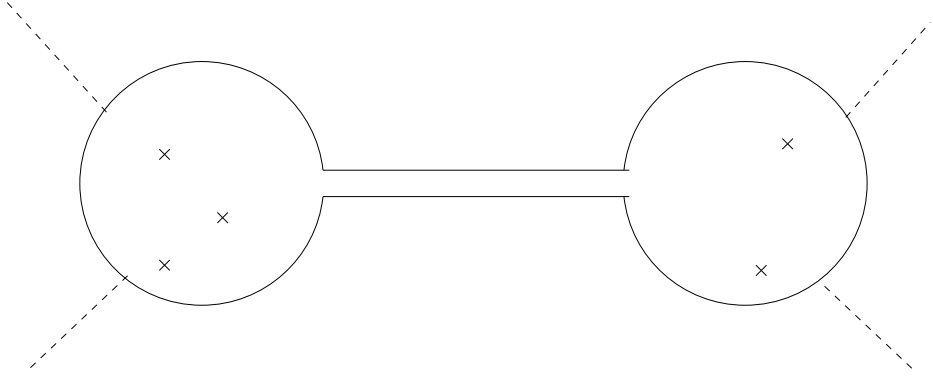}
  \caption{The pinching of the world-sheet in two halves.  The crosses mark
    the position of the world-sheet instantons.}
  \label{fig:world-sheet_factorisation}
\end{figure}

In Berkovits's string the amplitudes are correlation functions of vertex
operators in a background gauge field of world-sheet instantons.  In the
factorisation limit, the vertex operators (instantons) are constrained to
be on the border of (in the interior of) the left or right disks.  The
correlation function of the vertex operators is reduced to an integral over
the zero modes of the covariant derivatives in this world-sheet gauge
field, which are polynomial in the world-sheet variable $z$.  The integral
over the zero modes is then an integral over the coefficients of these
polynomials.  If the length of the strip is $L$, which we take to be very
large,\footnote{We will show below that we can make $L \rightarrow \infty$
  close to a multi-particle pole.  For now, we explore the consequences of
  this choice.} this implies some scaling with $L$ of these coefficients.

We will be interested in the factorisation of an amplitude with $n$ legs
and $d + 1$ negative helicity legs into two parts with $n_l$ and $n_r$ legs
($n_l + n_r = n$) and $d_l + 1$ and $d_r + 1$ negative helicity legs ($d_l
+ d_r = d$).

If we choose the coordinates such that zero is somewhere on the left,
(see Fig.~\ref{fig:world-sheet_factorisation}) the scaling with $L$ of
the moduli is such that
\begin{equation} a_k =
  \begin{cases}
    \hat{a}_{d_l-k} L^{d_r},& \quad \text{if $0 \leq k \leq
      d_l$},\\
    \bar{a}_{k-d_l} L^{d - k},& \quad \text{if $d_l \leq k \leq d$},
  \end{cases}
\end{equation} where $a_k$ is a generic modulus (may be bosonic or
fermionic).  Note that the two conditions above have an overlap for $k =
d_l$.  One consequence of this is $\hat{a}_0 = \bar{a}_0$.  The
hatted and barred variables should be considered of order zero in $L$.

For the positions of the vertex operators, we have
\begin{equation} \sigma_i =
  \begin{cases}
    \frac 1 {\hat{\sigma}_i},& \quad \text{if $i$ is on the
      left},\\
    \bar{\sigma}_i L,& \quad \text{if $i$ is on the right}.
  \end{cases}
\end{equation} Note that if some $\hat{\sigma} = 0$ or $\bar{\sigma} = 0$,
the scaling proposed above does not work.  This will become important in
the following.

If we also transform the $\xi_i$'s as
\begin{equation} \xi_i =
  \begin{cases}
    \hat{\xi}_i \hat{\sigma}_i^{d_l} L^{-d_r},& \quad \text{if $i$ is on
      the left},\\
    \bar{\xi}_i L^{-d} \bar{\sigma}_i^{-d_l},& \quad \text{if $i$ is on the
      right}.
  \end{cases}
\end{equation}

In terms of these variables, the zero modes
\begin{equation}
  \xi_i Z^I(\sigma_i) = \xi_i \sum_{k=0}^d a_k^I \sigma_i^k,
\end{equation} can be written as
\begin{equation}
  \xi_i Z^I(\sigma_i) = \hat{\xi}_i \sum_{k=0}^{d_l} \hat{a}_k^I
  \hat{\sigma}_i^k + \hat{\xi}_i \sum_{k=1}^{d_r} \bar{a}_k^I
  \hat{\sigma}_i^{-k} L^{-k},
\end{equation} if $i$ is on the left side, and
\begin{equation}
  \xi_i Z^I(\sigma_i) = \bar{\xi}_i \sum_{k=0}^{d_r} \bar{a}_k^I
  \bar{\sigma}_i^k + \bar{\xi}_i \sum_{k=1}^{d_l} \hat{a}_k^I
  \bar{\sigma}_i^{-k} L^{-k},
\end{equation} if $i$ is on the right side.

From these expressions it is obvious that, at the leading order in $L$, we
have a self-similar structure in the left and right sides.  If are allowed
to take the limit $L \rightarrow \infty$ the variables of integration
($a_k$, $\sigma_i$ and $\xi_i$) almost factorise.  Almost, because we still
have $\bar{a}_0 = \hat{a}_0$.

Does the limit $L \rightarrow \infty$ correspond to an internal line going
on shell?  To see that it does, integrate the moduli corresponding to the
$\mu$ to get equations for the $\tilde{\lambda}$,
\begin{align}
  \sum_{i \in L} \hat{\xi}_i \hat{\sigma}_i^k
  \tilde{\lambda}_i^{\dot{\alpha}} + \sum_{i \in R} \bar{\xi}_i
  \bar{\sigma}_i^{d_l-k} L^{-k} \tilde{\lambda}_i^{\dot{\alpha}} =& 0,
  \qquad & \text{for $d_l \geq k > 0$},\\ \sum_{i \in L} \hat{\xi}_i
  \tilde{\lambda}_i^{\dot{\alpha}} + \sum_{i \in R} \bar{\xi}_i
  \tilde{\lambda}_i^{\dot{\alpha}} =& 0,\\ \sum_{i \in L} \hat{\xi}_i
  L^{-k} \bar{\sigma}_i^{-k} \tilde{\lambda}_i^{\dot{\alpha}} + \sum_{i \in
    R} \bar{\xi}_i \bar{\sigma}_i^k \tilde{\lambda}_i^{\dot{\alpha}} =& 0,
  \qquad & \text{for $d_r \geq k > 0$}.
\end{align}

Combining these with the formulae for $\lambda$ yields
\begin{equation}
  P^{\alpha \dot{\alpha}} = \sum_{i \in L} \lambda_i^\alpha
  \tilde{\lambda}_i^{\dot{\alpha}} = \hat{a}_0^\alpha \sum_{i \in L}
  \hat{\xi}_i \tilde{\lambda}_i^{\dot{\alpha}} + \sum_{i \in L}
  \sum_{k=1}^{d_r} \hat{\xi}_i \bar{a}_k^\alpha L^{-k} \hat{\sigma}_i^{-k}
  \tilde{\lambda}_i^{\dot{\alpha}} - \sum_{i \in R} \sum_{k=1}^{d_l}
  \bar{\xi}_i \hat{a}_k^\alpha \bar{\sigma}_i^{d_l-k} L^{-k}
  \tilde{\lambda}_i^{\dot{\alpha}},
\end{equation} which in the limit $L \rightarrow \infty$, reduces to
\begin{equation}
  P^{\alpha \dot{\alpha}} \rightarrow \hat{a}_0^\alpha \sum_{i \in L}
  \hat{\xi}_i \tilde{\lambda}_i^{\dot{\alpha}},
\end{equation} which is on shell.  We also have
\begin{equation}
  P^2 = \frac 1 L \sum_{i, j \in L} \hat{\xi}_i \hat{\xi}_j
  [i\ j] \langle \hat{a}_0\ \bar{a}_1\rangle \frac 1 {\hat{\sigma}_j} + \frac
  1 L \sum_{i, j \in R} \bar{\xi}_i \bar{\xi}_j [i\ j] \langle \hat{a}_0\
  \hat{a}_1 \rangle \bar{\sigma}_j^{d_l-1} + \mathcal{O}(L^{-2}).
\end{equation} This means that the limit $L \rightarrow \infty$ corresponds
to the limit $P^2 \rightarrow 0$, as expected.

Another important thing to notice is the way the `gauge' symmetry
$\text{Gl}(2)$ acts on the hatted and barred variables.  The action on
$\hat{\sigma}$ and $\bar{\sigma}$ is easy to find.  We have
\begin{gather}
  \hat{\sigma} \rightarrow \frac {\delta \hat{\sigma} + \gamma}{\beta
    \hat{\sigma} + \alpha},\\
  \bar{\sigma} \rightarrow \frac {\alpha \bar{\sigma} + \frac \beta
    L}{\gamma L \bar{\sigma} + \delta}.
\end{gather}

The $\text{Gl}(2)$ action on the moduli $a_k$ and on the $\xi_i$ is more
complicated but can be obtained by using the action on the $\sigma_i$ along
with the invariance of $Z(\sigma_i)$.  What is somewhat remarkable is that
the hatted and barred variables inherit the same kind of $\text{Gl}(2)$
action as the initial variables.  More precisely, at the leading order in
$L$, the link between the action on the $\hat{\sigma}$ (or $\bar{\sigma}$)
and the action on $\hat{a}$ and $\hat{\xi}$ (or $\bar{a}$ and $\bar{\xi}$)
is the same as the link between the action on $\sigma$ and the action on
$a$ and $\xi$. (This conclusion is trivially true for the variables on the
left-hand side, but not for the variables on the right-hand side.  In fact,
$\bar{\xi}$ transforms as expected modulo an irrelevant multiplicative
factor.)

We can regard this passage from the initial integration variables to the
new ones (the barred and the hatted variables) as a change of variables in
the integrals forming the amplitude.  More precisely, it should be
considered as change of variables in a sub-domain of the full integration
domain where the scaling of the integration variables with $L$ is as
considered above.  This excludes for example the possibility that
$\hat{\sigma}$ or $\bar{\sigma}$ be zero; in fact, these conditions exclude
an open set around zero.  Since our purpose here is to display the
factorisation on a multi-particle pole, restricting the integration domain
need not bother us as long as this integration domain contains all the
contributions to this multi-particle pole.

This change of variables for the moduli $a$ has Jacobian one because the
bosonic and fermionic contributions cancel, the $\xi$ integrals are
homogeneous and do not contribute.  The only contribution comes from the
$\sigma$s.

Now the product of $\sigma$s can be written\footnote{We have here
  introduced a $0$, anticipating the fact that the internal line will be
  attached at $\sigma=0$.  See below.}
\begin{multline}
  \prod_{i=1}^n \frac {d \sigma_i}{\sigma_i - \sigma_{i + 1}} = \frac
  {\prod_{i=1}^{n_l} d \sigma_i} {(\sigma_1 - \sigma_2) \cdots
    (\sigma_{n_l} - 0) (0 - \sigma_1)} \\ \times \frac {\prod_{i=n_l+1}^n d
    \sigma_i} {(\sigma_{n_l + 1} - \sigma_{n_l + 2}) \cdots (\sigma_n - 0)
    (0 - \sigma_{n_l + 1})} \times \frac {\sigma_{n_l} (- \sigma_1)
    \sigma_n (- \sigma_{n_l + 1})}{(\sigma_{n_l} - \sigma_{n_l + 1})
    (\sigma_n - \sigma_1)} =\\= \frac 1 L \times \frac {\prod_{i=1}^{n_l} d
    \hat{\sigma}_i} {(\hat{\sigma}_1 - \hat{\sigma}_2) \cdots} \times \frac
  {\prod_{i=n_l+1}^n d \bar{\sigma}_i}{(\bar{\sigma}_{n_l + 1} -
    \bar{\sigma}_{n_l + 2}) \cdots} \left(1 + \mathcal{O}(L^{-1})\right).
\end{multline} Therefore, this part also factorises at the leading order in
$L$.

In order to prove factorisation, we need to introduce two more vertex
operators corresponding to the internal line going on-shell in the
factorisation limit.  Use the notation
\begin{equation}
  \Psi_{\pi, \bar{\pi}, \eta}(\lambda, \mu, \psi) = \int
  \frac {d \xi} \xi \delta^2(\pi - \xi \lambda) \exp(i \xi \left[\mu,
    \bar{\pi}\right]) \exp(i \xi \psi^A \eta_A).
\end{equation}

For these wavefunctions we can prove orthonormality and completeness
relations.

Orthonormality\footnote{The factor $(2 \pi)^2$ that seems to be missing has
been included in the integration measure.}:
\begin{multline}
  \int \frac {d^2 \lambda d^2\mu d^4 \psi}{\text{Gl}(1)} \Psi^*_{\pi,
    \bar{\pi}, \eta}(\lambda, \mu, \psi) \Psi_{\pi', \bar{\pi}',
    \eta'}(\lambda, \mu, \psi) =\\= \int \frac {d \xi} \xi \delta^2(\xi \pi
  - \pi') \delta^2(\bar{\pi} - \xi \bar{\pi}') \delta^4(\eta - \xi \eta')
  \equiv \delta_{\pi, \bar{\pi}, \eta; \pi', \bar{\pi}', \eta'}.
\end{multline} The $\text{Gl}(1)$ group comes from the following symmetry
of the integral
\begin{align}
  \lambda \rightarrow & t \lambda,\\
  \mu \rightarrow & t \mu,\\
  \psi \rightarrow & t \psi,\\
  (\xi, \xi') \rightarrow & t^{-1} (\xi, \xi').
\end{align}

Completeness:
\begin{multline}
  \int \frac {d^2 \pi d^2 \bar{\pi} d^4 \eta}{\text{Gl}(1)} \Psi^*_{\pi,
    \bar{\pi}, \eta}(\lambda, \mu, \psi) \Psi_{\pi, \bar{\pi},
    \eta}(\lambda', \mu', \psi') =\\= \int \frac {d \xi} \xi
  \delta^2(\lambda - \xi \lambda') \delta^2(\mu - \xi \mu') \delta^4(\psi -
  \xi \psi') \equiv \delta_{\lambda, \mu, \psi; \lambda', \mu', \psi'}.
\end{multline} The $\text{Gl}(1)$ group comes from the following symmetry
of the integral
\begin{align}
  \pi \rightarrow & t \pi,\\
  \bar{\pi} \rightarrow & t^{-1} \bar{\pi},\\
  \eta \rightarrow & t^{-1} \eta,\\
  (\xi, \xi') \rightarrow & t (\xi, \xi').
\end{align} The measure transforms like
\begin{equation}
  d^2 \pi d^2 \bar{\pi} d^4{\eta} \rightarrow t^4 d^2 \pi d^2 \bar{\pi}
  d^4{\eta}
\end{equation} and a factor in the integrand
\begin{equation}
  \delta^2(\pi - \xi \lambda) \delta^2(\pi - \xi' \lambda')
  \rightarrow t^{-4} \delta^2(\pi - \xi \lambda) \delta^2(\pi - \xi'
  \lambda').
\end{equation} (These combine to render the integral invariant.)

We use the above formulae to separate the integrals over moduli (remember
that we still have the constraint $\hat{a}_0 = \bar{a}_0$).  Then,
schematically
\begin{multline}
  \int d^{4(d+1)|4(d+1)} a \cdots = \int d^{4(d_l+1)|4(d_l+1)} \hat{a}
  d^{4(d_r+1)|4(d_r+1)} \bar{a} \delta^{4|4}(\hat{a}_0 - \bar{a}_0)
  \cdots =\\= \int \frac {d^2 \pi d^2 \bar{\pi} d^4
    \eta}{\text{Gl}(1)} \int \frac {d^{4(d_l+1)|4(d_l+1)} \hat{a}
    d^{4(d_r+1)|4(d_r+1)} \bar{a}} {\text{Gl}(1)} \Psi^*_{\pi,
    \bar{\pi}, \eta}(\hat{a}_0) \Psi_{\pi, \bar{\pi},
    \eta}(\bar{a}_0) \cdots,
\end{multline} where we have inserted the delta function from the
completeness relation and $\text{Gl}(1)$ acts projectively on $\bar{a}$ or
$\hat{a}$ moduli.  The dots in the above formula stand for a function of
the moduli $a$ which is invariant under a scaling of $\bar{a}$ and
$\hat{a}$ separately.  At the dominant order in $L$ this property is
satisfied by the integrands we consider.  The delta function really stands
for $\delta^{4|4}(\hat{Z}(\sigma)-\bar{Z}(\sigma'))$, understood to be
evaluated at $\sigma = \sigma' = 0$.

\subsection{Gauge-fixing}

Now we come to the issue of gauge-fixing.  We can fix the gauge in several
different ways.  If we gauge-fix one component of $\hat{a}_0$, say
$\hat{a}_0^1$, $\hat{\sigma}_i$, $\hat{\sigma}_j$ and $\bar{\sigma}_p$ we
get a Jacobian
\begin{equation}
  \label{eq:Jacobian_J}
  J = - \frac 1 L \hat{a}_0^1 (\hat{\sigma}_i - \hat{\sigma}_j)(-1 + L
  \bar{\sigma}_p \hat{\sigma}_i)(-1 + L \bar{\sigma}_p \hat{\sigma}_j).
\end{equation} The $\text{Gl}(1)$ gauge invariance at the right can be
gauge-fixed independently and gives a Jacobian $\bar{a}_0^1$, for example.
Note that if anyone of $\hat{\sigma}_i$, $\hat{\sigma}_j$ or
$\bar{\sigma}_p$ is zero, the Jacobian is $J \sim L^{-1}$ and this will not
contribute in the factorisation limit (note that in order for a
contribution to contribute in the limit $L \rightarrow \infty$ it has to
cancel the factor in $\frac 1 L$ from the product of $\sigma$'s).  This is
consistent with the interpretation we gave that the internal line has
$\hat{\sigma} = \bar{\sigma} =0$ so, in same sense it already is
gauge-fixed at zero.

Alternatively, we could try to gauge-fix three $\hat{\sigma}$'s and a
modulus on the left-hand side which would seem like over-fixing once the
internal line is taken into account.  However, this gauge-fixing gives a
Jacobian of order $L^0$, and thus doesn't contribute in the
factorisation limit.

When $\hat{\sigma}_i$, $\hat{\sigma}_j$ and $\bar{\sigma}_p$ are different
from zero we have
\begin{equation}
  J = - L \hat{a}_0^1 \bar{\sigma}_p^2 \hat{\sigma}_i
  \hat{\sigma}_j (\hat{\sigma}_i - \hat{\sigma}_j) + \mathcal{O}(1).
\end{equation} We want to compare this with the case when the left- and
right-hand sides are completely gauge-fixed, however on the right-hand side
there are only two $\bar{\sigma}$'s and a modulus fixed.  We use the fact
that $\bar{a}_0^1 \bar{\sigma}_p \bar{\sigma}_q (\bar{\sigma}_p -
\bar{\sigma}_q)$ times the right hand integrals where we don't integrate
over $\bar{a}_0^1$, $\bar{\sigma}_p$, $\bar{\sigma}_q$ is independent of
$\bar{\sigma}_q$ and can be taken out of the integral over
$\bar{\sigma}_q$.

Dividing the full Jacobian $J$ by the Jacobians needed to recombine the
left and right parts into gauge invariant amplitudes gives
\begin{equation}
  \frac J {L J_l J_r} = \frac {\bar{\sigma}_p}{\bar{\sigma}_q
    (\bar{\sigma}_p - \bar{\sigma}_q)} = \frac 1 {\bar{\sigma}_q} + \frac 1
  {\bar{\sigma}_p - \bar{\sigma}_q},
\end{equation} where $J_l = \hat{\sigma}_i \hat{\sigma}_j (\hat{\sigma}_i -
\hat{\sigma}_j)$ and $J_r = \bar{\sigma}_p \bar{\sigma}_q (\bar{\sigma}_p -
\bar{\sigma}_q)$.  We are left with the integral over $\bar{\sigma}_q$
\begin{equation}
  \oint d \bar{\sigma}_q \left(\frac 1 {\bar{\sigma}_q} -
    \frac 1 {\bar{\sigma}_q - \bar{\sigma}_p}\right).
\end{equation} This integral is zero if we interpret it in the most naive
way possible, by taking a contour around $0$ and $\bar{\sigma}_p$ in the
$\bar{\sigma}_q$ plane.  However, we have to recall that the region in the
neighbourhood of $\bar{\sigma}_q = 0$ is special and is not included in our
integration domain (at any rate, a contour around zero which is included in
our integration domain cannot be shrunk to $\bar{\sigma} = 0$ while
staying inside the integration domain).  Therefore, we propose to do the
next less naive thing possible and take a contour which does not go around
$\bar{\sigma} = 0$.  The result of the integration is then $-2 \pi i$.

\subsection{Testing the Contour Prescription}

Since our prescription for the choice of contour is not very well
justified, it is important to test it by using different gauge fixing
conditions.  We test this by gauge-fixing the linear combination
$\bar{\sigma}_p + \zeta \bar{\sigma}_q$.  This has a Jacobian $J_\zeta$
which is such that
\begin{equation}
  \frac {J_\zeta}{L J_l J_r} = \frac 1 {\bar{\sigma}_q} -
  \frac {1 + \zeta}{\bar{\sigma}_p - \bar{\sigma}_q} - \frac \zeta
  {\bar{\sigma}_p} + \mathcal{O}(L^{-1}).
\end{equation} Now suppose $\bar{\sigma}_p + \zeta \bar{\sigma}_q$ is
gauge-fixed to a value $\tau$.  This is implemented by introducing a delta
function $\delta(\bar{\sigma}_p + \zeta \bar{\sigma}_q - \tau)$ and the
Jacobian $J_\zeta$ in the integral.  After integrating over
$\bar{\sigma}_p$ we are left with the following integral over
$\bar{\sigma}_q$
\begin{equation}
  \oint d \bar{\sigma}_q \left(\frac 1 {\bar{\sigma}_q} -
    \frac 1 {\bar{\sigma}_q - \frac \tau {1 + \zeta}} + \frac \zeta {\zeta
      \bar{\sigma}_q - \tau}\right).
\end{equation} In the integrand, the first and the last term correspond to
$\bar{\sigma}_q = 0$ and $\bar{\sigma}_p = 0$ respectively.  Therefore, as
before, we argue that the choice of contour is such that they don't
contribute.  The remaining term yields $-2 \pi i$.

There is however a problem for $\zeta = -1$ and, in this particular case,
our prescription does not work.  It does work however, for the whole family
of gauge-fixing conditions where $\zeta \neq -1$.  (Note that when $\zeta
\rightarrow -1$ the pole which contributes to the integral is sent to
infinity and also out of our domain of integration.)

\subsection{The $\frac 1 {P^2}$ Pole}

We still have the integration $\int \frac {d^2 \pi d^2 \bar{\pi} d^4
\eta}{\text{Gl}(1)}$ to perform.  Concentrate on the bosonic part.  After
gauge-fixing $\pi^1$ the measure is $\pi^1 d \pi^2 d^2 \bar{\pi}$.  The
right and left part each contain a momentum conserving delta function.
Denote $P$ the total momentum at left and $Q$ the total momentum at right
\begin{multline}
  \int \pi^1 d \pi^2 d^2 \bar{\pi}^{\dot{\alpha}} \delta^4(P^{\alpha
    \dot{\alpha}} - \pi^\alpha \bar{\pi}^{\dot{\alpha}}) \delta^4(Q^{\alpha
    \dot{\alpha}} - \pi^\alpha \bar{\pi}^{\dot{\alpha}}) =\\= \delta^4(P -
  Q) \int \pi^1 d \pi^2 d^2 \bar{\pi}^{\dot{\alpha}} \delta^4(P^{\alpha
    \dot{\alpha}} - \pi^\alpha \bar{\pi}^{\dot{\alpha}}).
\end{multline} The integral above can be computed straightforwardly and the
result is $\delta(P^2)$.

What we want to do now is to interpret this as a holomorphic delta function
$\delta(z) \equiv \frac 1 {2 \pi i} \frac 1 z$.  Note that this is not an
arbitrary assumption.  The delta functions from the connected prescription
are \emph{not} real delta functions, but they can be interpreted as complex
delta functions (see~\ref{sec:delta}). Granted this interpretation, we
obtain the pole we were looking for
\begin{equation}
  \delta(P^2) \equiv \frac 1 {2 \pi i} \frac 1 {P^2}.
\end{equation}

\subsection{The Remaining Ingredients}

\begin{itemize}
\item The fermionic integrals: It is easy to see that the integral
  over $\eta$ imposes the constraint that the sum of helicities at the
  two ends of the internal line be zero.  One might be worried that
  this allows the exchange of fermions and scalars also but this is
  not so since the amplitudes with one fermion/scalar and all the rest
  vector particles vanish.  Therefore, the only contribution which
  survives when taking the residue is the exchange of an internal
  gluon.
\item The gauge coupling: it's $g^{n-2}$ for the initial amplitude and
  $n=n_l+n_r$.  Upon factorisation we can write this as $g^{(n_l+1)-2}
  \times g^{(n_r+1)-2}$ which is the correct coupling factor.
\item The colour factors: They factorise as shown
  in~\cite[eq.~6.20]{Mangano:1990by}.  For $SU(N)$ groups, there is a
  sub-dominant factor in $\tfrac 1 N$ but it vanishes at the pole.
\end{itemize}

\section{Summary and Discussion}
\label{sec:discussion}

In this paper we have given some arguments supporting the factorisation of
the connected prescription for Yang-Mills amplitudes.  This is not a
complete proof, however since the choice of contours is not well
understood.  This is not a new problem.  It is already implicit in
refs.~\cite{Roiban:2004vt,Roiban:2004ka,Roiban:2004yf}, and also shows up
in ambiguities in the contour deformation argument of
ref.~\cite{Gukov:2004ei}.

In~\cite{Roiban:2004yf}, this problem was avoided by summing over all
solutions at finite vertex positions and finite moduli, so the
contours had to be specified only vaguely by saying that they have to
encircle all possible singularities.  Here we have followed the same
strategy, adding the further constraint of staying inside a specified
domain.

It seems to us that there may be several prescriptions for the
integration contours which give the same results as far as tree
amplitudes are concerned.  Berkovits's model seems to require an
ordering of the vertex operators $\sigma_i < \sigma_{i+1}$ on the
border of the disk, whereas the moduli are unrestricted.  The
connected prescription, on the other side, doesn't seem to impose such
ordering restrictions on the $\sigma$s but could impose some
restriction on the moduli in keeping with our interpretation of the
delta functions (see our discussion above about the contour from zero
to infinity and also the discussion in~\cite{Cachazo:2004kj}).  It
might very well be that the contours in one can case can be deformed
to the contours in the other case, but a priori they are different.
One may wonder whether either of these contour prescriptions can be
extended consistently to loop level.  Of course, one first needs a
better understanding of the contours at tree level.  This would also
enable a study of corrections to the strict $L=\infty$ limit
(corresponding to corrections to $C=0$ in ref.~\cite{Gukov:2004ei}),
yielding a quantitative computation of the pole and its residue.

It would also be interesting to see what the arguments presented
in this paper have to say about factorisation in the case of conformal
supergravity~\cite{Berkovits:2004jj} or Einstein
supergravity~\cite{Abou-Zeid:2006wu}.  Most of the discussion carries
over; only the wavefunctions need to be modified

\begin{acknowledgments}
I thank I.~Bena, D.~Kosower, L.~Motl and R.~Roiban for discussions and
encouragement.  I am especially indebted to D.~Kosower and R.~Roiban for
their critical comments.
\end{acknowledgments}

\end{document}